\documentclass[aps,prb,twocolumn,superscriptaddress,showpacs]{revtex4}%
\usepackage{amssymb}
\usepackage{amsmath}
\usepackage{graphicx}
\usepackage{amsfonts}%
\usepackage{subfigure}
\begin{document}
\title{Potential and current distribution in strongly anisotropic Bi$_{2}$Sr$_{2}
$CaCu$_{2}$O$_{8}$ single crystals at current breakdown}
\author{I. Pethes}
\affiliation{Research Institute for Solid State Physics and Optics, PO Box 49, H-1525
Budapest, Hungary}
\author{A. Pomar\footnote{Present address: Institut de Ci\`encia de
Materiales de Barcelona, CSIC, Compus de la UAB, 08193 Barcelona, Spain}}
\affiliation{Service de Physique de l'Etat Condens\'{e}, Commissariat \`{a} l'Energie
Atomique, Saclay, F-91191 Gif-sur-Yvette, France}
\author{B. Sas}
\author{G. Kriza}
\affiliation{Research Institute for Solid State Physics and Optics, PO Box 49, H-1525
Budapest, Hungary}
\author{K. Vad}
\affiliation{Institute of Nuclear Research, PO Box 51, H-4001 Debrecen, Hungary}
\author{\'{A}. Pallinger}
\affiliation{Research Institute for Solid State Physics and Optics, PO Box 49, H-1525
Budapest, Hungary}
\author{F. Portier}
\author{F. I. B. Williams}
\affiliation{Service de Physique de l'Etat Condens\'{e}, Commissariat \`{a} l'Energie
Atomique, Saclay, F-91191 Gif-sur-Yvette, France}
\date{\today}

\begin{abstract}
Experiments on potential differences in the low-temperature vortex solid phase of
monocrystalline platelets of superconducting $\mathrm{Bi_{2}Sr_{2}%
CaCu_{2}O_{8}}$ subjected to currents driven either through an \textit{ab}
surface or from one such surface to another show evidence of a resistive/nonresistive front moving progressively out from the current contacts as the
current increases. The depth of the resistive region has been measured by
an in-depth voltage probe contact. The position of the front associated
with an injection point appears to depend only on the current magnitude and
not on its withdrawal point. It is argued that enhanced nonresistive
superconducting anisotropy limits current penetration to depths less than the London
length and results in a flat rectangular resistive region with simultaneous
$ab$ and $c$ current breakdown which moves progressively out from the
injection point with increasing current. Measurements in \textit{ab} or \textit{c}
configurations are seen to give the same information, involving both
$ab$-plane and $c$-axis conduction properties.

\end{abstract}


\maketitle

\section{Introduction}

Understanding the current and the potential distributions at the onset of
dissipation in a superconductor is important; at a fundamental level for
interpreting the results of transport measurements in terms of the
force-velocity relation for the vortices and the Josephson coupling 
between layers as well as on a practical level
for understanding how to maximize the current carrying capacity of a wire. The
family of high-$T_{c}$ cuprate superconductors naturally raises the question
of the influence of high anisotropy on current carrying properties. How does
one interpolate between the extreme anisotropic limit, where current injected
into a superconducting plane has no transfer to other planes, and the isotropic
case where the penetration of the current in the superconducting state is
determined by the London screening length.\cite{Londonbook} To give a concrete
illustration of the problem, a naive interpretation of the threshold current
for dissipation in a strongly anisotropic $\mathrm{Bi_{2}Sr_{2}CaCu_{2}O_{8}}$
(BSCCO) single crystal based on penetration limited to the London screening
length or, even more naively, supposed to be uniform over the typical
thickness of a few micrometers gives threshold current densities one to two
orders of magnitude lower than a standard interpretation of a magnetic
hysteresis loop.

Anisotropy is a central feature of the family of the high-$T_{c}$
cuprate superconductors which are usually modeled by discrete superconducting
sheets parallel to the $ab$ crystallographic plane weakly coupled together in
the $c$ direction by extended Josephson junctions.\cite{lawrence1971} The weak
coupling between planes gives rise to very high conduction anisotropy in both
the normal and superconducting states. It is also responsible for the richness
of the phase diagram of the vortices created when one applies a magnetic
field. A convenient experimental probe for investigating these phases,
particularly as regards their pinning to the host lattice disorder, is to look
at the voltage response to transport current ($VI$ characteristics) for currents which take one into the resistive regime
where the vortices are dislodged. But to interpret these measurements
correctly, it is important to know about the current distribution in the
resistanceless regime, how it is modified as dissipation sets in and of course
what the potential distribution then looks like.

It was remarked some time ago,\cite{busch1992} in connection with
resistivity experiments, that if the response of the
superconductor is Ohmic, {\boldmath$\mathbf{E=\rho\cdot J}$} (the resistivity
tensor {\boldmath$\mathbf{\rho}$} is constant and independent of current and
position), the problem may be treated as for the normal state where it is
implicit that the phase slip giving rise to the voltage also relaxes the
magnetic dephasing to allow the current to penetrate beyond the London
screening depth. In this regime, at least, the high
anisotropy of the (BSCCO) samples studied limited the current penetration
into the depth of the sample to a small fraction of its thickness
and that it would therefore be quite
erroneous to deduce a critical current density on the basis of uniform
penetration across the section of a short sample. Experiments to date on 
the current profile have been able to show a concentration of current 
towards the sample edges, but have been insensitive to the depth 
dependence\cite{FuchsNature} which is the aspect that interests us chiefly here.

\begin{figure}[ptb]
\includegraphics[width=8.6cm]{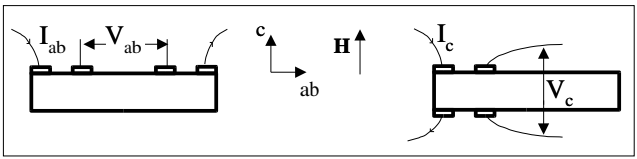}\caption{$ab$ and $c$ contact
configurations.}%
\label{fig:configurations}%
\end{figure}

The present paper reports an experimental investigation and a few basic
reflections on this problem in the non-Ohmic, low-temperature, high magnetic-field vortex solid phase, also in monocrystalline BSCCO in a \textit{c}-directed 
magnetic field. The experiment consists of measuring the voltage response to short 
pulses of current up to and beyond dissipative breakdown. Potentials are measured 
at contacts placed in the 
usual way on the \textit{ab} faces supplemented by in-depth potential 
contacts to have direct access to depth-dependent features. One of the important 
points that emerges is that the physically
unrelated $c$-axis and $ab$-plane dissipative breakdowns interact to create a
resistive/nonresistive front which has the effect of altering the current
distribution and giving rise to a critical current which involves both $ab$
and $c$ characteristics. Not only does the voltage response along an
$ab$-plane surface into which the current is injected and withdrawn (the
``$ab$ configuration'' of Fig.\ \ref{fig:configurations}) show features of the
Josephson junctions between planes, but also the response to current injection
and withdrawal on opposing $ab$-plane faces (the ``$c$ configuration'' of
Fig.\ \ref{fig:configurations}) shows features of breakdown in the $ab$ plane.
One manifestation of this is that the threshold current in the $ab$
configuration on BSCCO single crystals\cite{Sas2000} shows the same
temperature and preparation dependence as for the $c$
configuration.\cite{delacruz} This brings up the question as to who
measures what and does one measure what one thinks; but beyond that what do
the results mean?

\section{Experimental technique}

Our samples are thin monocrystalline platelets with the $c$ axis perpendicular
to the face. As all the samples studied gave qualitatively the same results
for similar contact configurations, we limit ourselves here to the results on
one sample of each different contact configuration. The sizes of samples $A$,
$B$, and $C$ are about $2\times0.35\times0.005$~$\mathrm{\ mm^{3}}$,
$1.1\times0.3\times0.002$~$\mathrm{mm^{3}}$, and $0.7\times0.5\times
0.005$~$\mathrm{mm^{3}}$. They were all fabricated by a melt cooling
technique.\cite{keszei1989, cooper1990} The critical temperatures were between
88 and 90~K with a transition width of about 2~K at zero field. The anisotropy
coefficient $\gamma_{n}\approx500$ was estimated from normal-state
resistivities with $\rho_{\text{ab}}\approx$~100~$\mu \Omega$ cm at 90~K. In the case of samples $A$ and $B$, electrical contact was made
by bonding 25-$\mu$m gold wires with silver epoxy fired at 900~K. For sample $C$ the contacts
were made by depositing silver on a lithographically defined area using a
lift-off technique and heat treating at 700~K for 1000~s. In all cases the
contact resistance was less than 3~$\Omega$. The sample was mounted flat against a 7-mm-diameter 0.5-mm-thick sapphire
disk with silicone grease to ensure thermal homogeneity and mechanical freedom
and placed in the bore of a superconducting magnet with the field along the
$c$ direction. The sample and disk were surrounded by exchange gas and the
temperature was electronically regulated. The longitudinal voltage-current 
($VI$) characteristics were measured by a symmetrized differential four-point
technique using 25-$\mu$s triangular pulses (12.5 $\mu\text{s}$
 from zero to maximum current) of maximum amplitude in the range
10~mA~$<I_{m}<$~350~mA, usually restricted to exceed the threshold by about
30\%, with repetition time $0.1$~s. As reported earlier,\cite{Sas2000} the
observed threshold is independent of pulse duration for times up to about 250
$\mu\text{s}$, argued there to be the time required for heat produced in the current
contact to diffuse towards the voltage contact, whereas bulk heating in the
superconductor was estimated not to be important. Further technical details
may be found in Ref. \onlinecite{Sas2000}.

\section{Experimental results}

We present first the results of experiments on the $VI$ response with surface
contacts in the standard $ab$ and $c$ configurations on the same sample using
contacts common to the two configurations. The results lead rather naturally
to a description in terms of resistive/nonresistive fronts moving outwards
from the current contacts. Experiments are then presented for other surface
contact configurations intended to check this hypothesis. These are followed
by experiments with a different type of contact configuration designed to probe
the depth of the sample as breakdown progresses.

\begin{figure}[ptb]
\begin{center}
\subfigure[]{
\includegraphics[width=7.6cm]{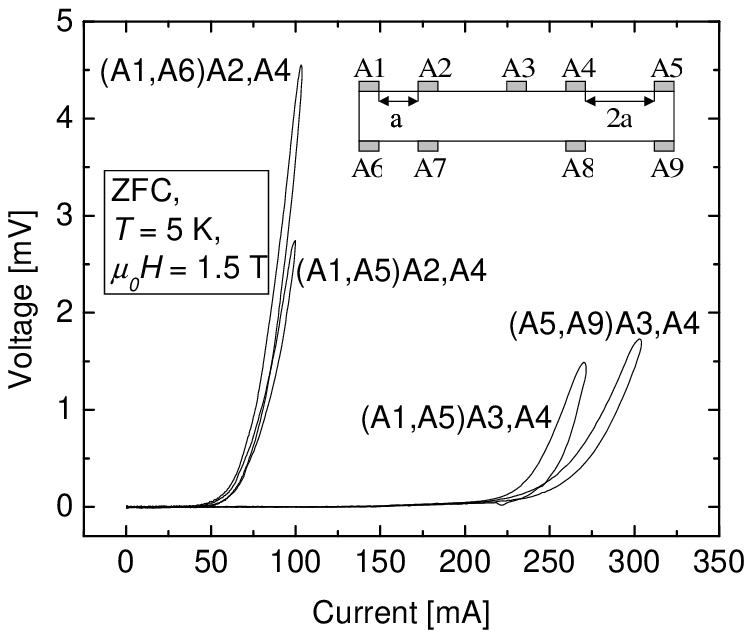}}
\subfigure[]{
\includegraphics[angle=-90,width=7.6cm]{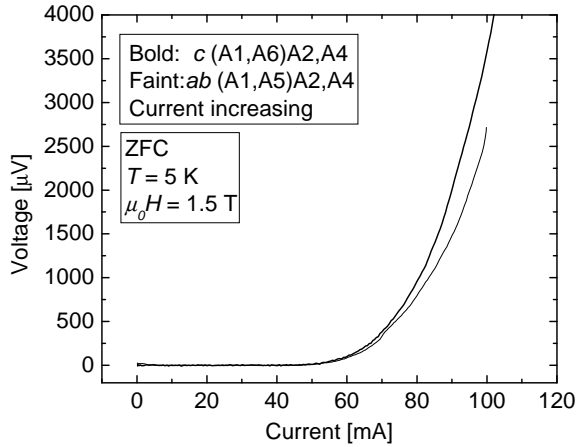}}
\subfigure[]{
\includegraphics[angle=-90,width=7.6cm]{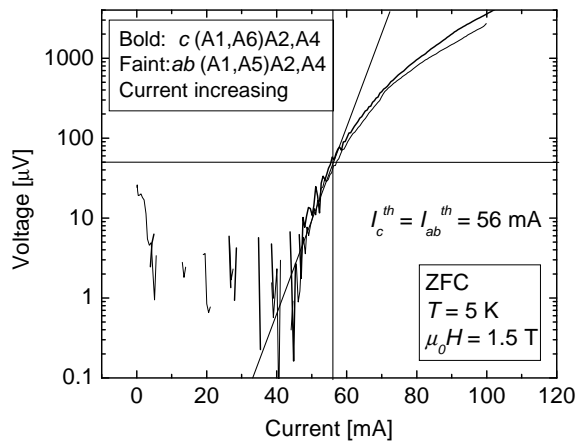}}
\caption{$VI$ curves measured on
sample $A$. The letters on the $VI$ curves refer to the contacts used: the
first pair in brackets denote the current contacts and the last two the
potential contacts. The inset in (a) shows the contact configuration.
Part (a) shows the response to up and down current sweeps of isoceles triangular shape of
total duration 25 $\mu\text{s}$. Parts (b) and (c) show in more 
detail, on both
linear and semilogarithmic plots, upsweep response for the first two curves of
part (a).
}%
\label{fig:frontab}%
\end{center}
\end{figure}

\begin{figure}[ptb]
\includegraphics[height=7cm, angle=-90]{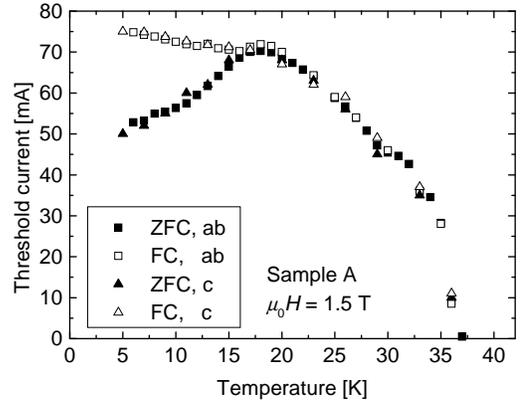}
\caption{Temperature
dependence of the threshold current with $ab$ and $c$ contact configurations
on the same sample.}
\label{fig:Ithtdep}
\end{figure}

\begin{figure}[ptb]
\includegraphics[height=7cm, angle=-90]{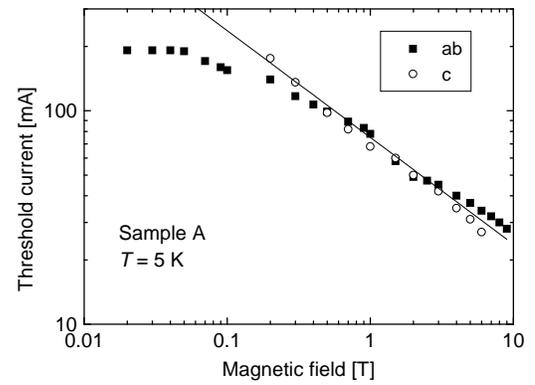}
\caption{Magnetic-field
dependence of the threshold current with both contact configurations on the
same sample. The solid line shows $H^{-1/2}$.}
\label{fig:Ithhdep}
\end{figure}

The first experiment was done on samples contacted to be able to measure in
both $ab$ and $c$ configurations on the same sample at the same time as
illustrated in Fig.\ \ref{fig:frontab}. The \textquotedblleft$ab$
configuration\textquotedblright\ measurements refer to the potential drop
between contacts ($A2,A4$) for current injected at $A1$ and withdrawn from $A5$.
The \textquotedblleft$c$ configuration\textquotedblright\ refers to the
potential measured across ($A2,A7$) for current passed through ($A1,A6$).
Figure \ref{fig:Ithtdep} shows the temperature dependence of the threshold
current in the $VI$ characteristic at $1.5$~T in sample $A$ for both
configurations. The threshold current is defined by the break in slope
criterion of Ref. \onlinecite{Sas2000} which gives essentially the
same values as the criterion of crossover from Kim-Anderson behavior at low
current to power law close to linear behavior at higher
current,\cite{portier2002} illustrated in detail for the first pair of curves
of Fig.\ \ref{fig:frontab} in the semilog plot in the same figure. The measured thresholds, as represented in
Fig.\ \ref{fig:Ithtdep}, are seen to be independent of the configuration,
whether the sample is field cooled (FC) or zero-field cooled (ZFC) prepared,
despite the fact that the two preparations give different thresholds for
temperatures lower than the peak in the ZFC results.\cite{Sas2000} For the FC
preparation, the field is applied above $T_{c}$ and the sample is cooled at
constant field to the lowest measuring temperature, subsequent measurements
being made on increasing the temperature at the same field. For ZFC
preparation, the system is cooled in zero field from above $T_{c}$ to the
lowest temperature before applying the field and then making measurements at
sequentially higher temperatures at the same field. Figure \ref{fig:Ithhdep}
shows the magnetic-field dependence in both contact configurations for ZFC
preparation at $5$~K on increasing the field (FC preparation followed by field
variation gives the same result, except for the initial FC prepared point
\cite{portier2002}). At fields above about $0.3$~T, the threshold currents for
the two different configurations again show the same behavior, this time in
field, varying approximately as $H^{-1/2}$. Not only are the temperature and
field dependences of the threshold currents in the two configurations similar
in form, but the values themselves coincide. Measurements along several other
lines in the $(H,T)$ plane confirm this indistinguishability of $ab$ and $c$
configuration thresholds as generic behavior for the low-temperature high-field domain. This fact led us to look in more detail with the following experiment.


Again referring to the inset of Fig.\ \ref{fig:frontab}, we used three
potential contacts ($A2,A3,A4$) on the top layer and two ($A7,A8$) on the
bottom. The contact $A2$ is half as far away from the left current contact
$A1$ as is the contact $A4$ from the right-hand current contact $A5$, while
$A3$ is midway between the two current contacts. If we apply the current
between contacts $A1$ and $A5$ ($ab$ direction) we measure a much lower
threshold current between $A2$ and $A4$ than between $A3$ and $A4$. On the
other hand, we find virtually identical $VI$ curves on the $ab$ potential
contacts ($A2,A4$) whether the current is applied in the $c$ direction through
contacts ($A1,A6$) or along the $ab$ direction ($A1,A5$); similarly the $VI$
across ($A3,A4$) is nearly identical for $ab$-directed current through
($A1,A5$) or for $c$-directed current through ($A5,A9$).

\begin{figure}[ptb]
\includegraphics[width=7cm]{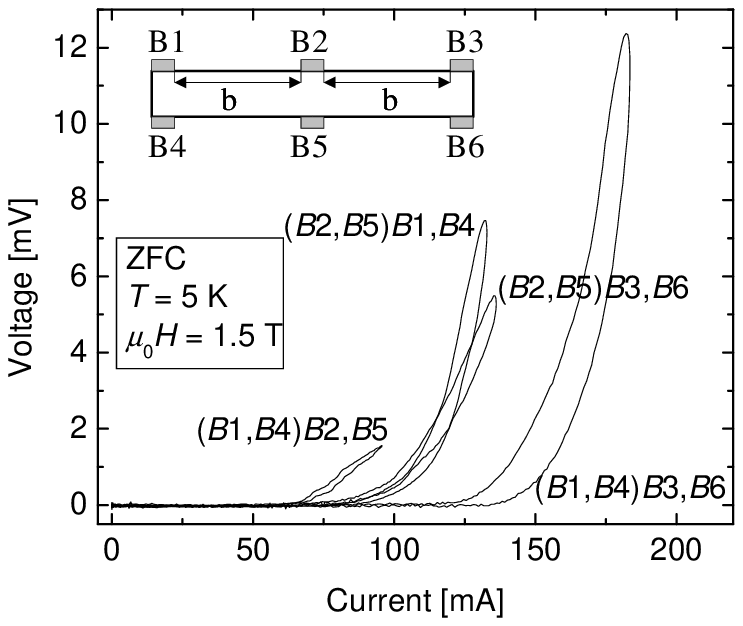}
\caption{$VI$ response of sample
$B$. The labeling convention is as before.}
\label{fig:frontc}
\end{figure}

These results suggest the idea that the sample first becomes resistive around
the injection points and that this resistive region progressively invades the
sample as the current is increased. This led us to experiment with the contact
configuration of sample $B$ illustrated in Fig.\ \ref{fig:frontc} where we
drove the current through the $c$ direction and also measured the voltage drop
across that direction. On sending the current through ($B1,B4$) the $VI$
characteristic measured across ($B2,B5$) shows a threshold current which is
about half that measured across contacts ($B3,B6$). On the other hand, if we
put the current through the middle contacts ($B2,B5$) we find the same
threshold at either of the two end potential contacts ($B1,B4$) or ($B3,B6$)
with a value intermediate between the previous two.

\begin{figure}[ptb]
\includegraphics[width=7cm]{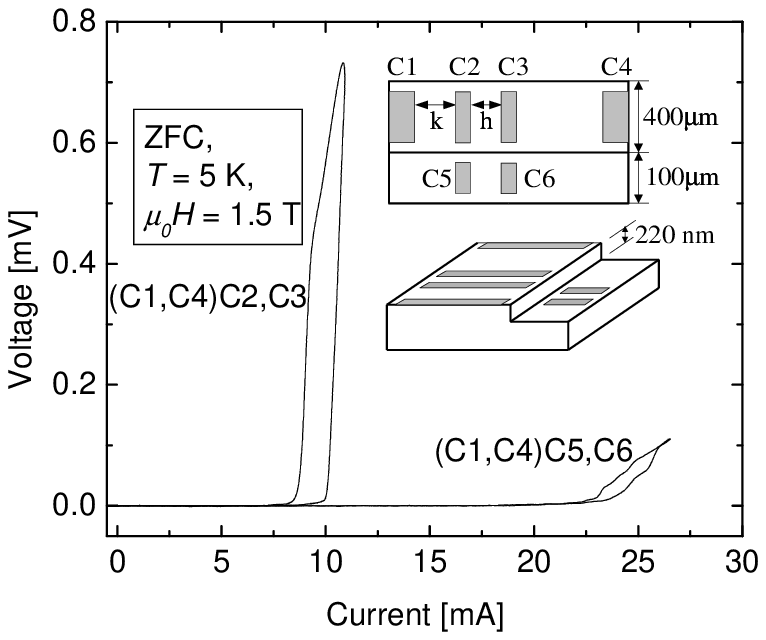}
\caption{$VI$ response for sample
$C$. The dimensions $k$ and $h$ are 125~$\mathrm{\mu}$m and 75~$\mathrm{\mu}$m,
respectively. The potential contact dimension along the sample length is
50~$\mathrm{\mu}$m.}%
\label{fig:depth}%
\end{figure}

Although these experiments lend considerable support to the idea of a
resistive/nonresistive front progressing along the surface, they say nothing
about its profile with depth since at no point were we able, in this 
low-temperature high-field phase and despite micron thin samples, to drive the
lower face contacts resistive with current sent along the top face. To have
more direct information about the shape of the front with depth, we prepared a
third sample $C$ as in Fig.\ \ref{fig:depth} with a lithographically defined
terrace argon ion etched along the length of one $ab $ face to a depth of
$220$~nm. The upper part of the face was also argon ion etched to avoid any
difference in surface pinning between it and the terrace, the width of which
was restricted to a small fraction ($20\%$) of the total sample width to
minimize the trivial effect of current spread as the distribution attains the
depth of the terrace. The contacts were accurately aligned between the two
levels by a scanning electron-beam masking microscope. On a control sample
with the same contact configuration but with no terrace, the $VI$ response on
contacts ($C2,C3$) gave the same values of the threshold current and dynamic
resistance as contacts ($C5,C6$), themselves very similar to those for the top
contacts of the terraced sample.

Sending an $ab$-directed current through contacts ($C1,C4$) on the
upper part of the face, we measured the potential drop across ($C2,C3$) on the
upper part and across ($C5,C6$) on the step. The threshold current for the
terrace contacts is considerably higher than for the corresponding top
contacts: about $25$~mA and $10$~mA, respectively, much greater than the
decrease of $20\%$ in the current density to be expected from simple current
spread into the wider part of the sample. We interpret the large difference to
arise from the current distribution with depth.

\section{Interpretation of experimental results}

\begin{figure}[ptb]
\includegraphics[width=8.6cm]{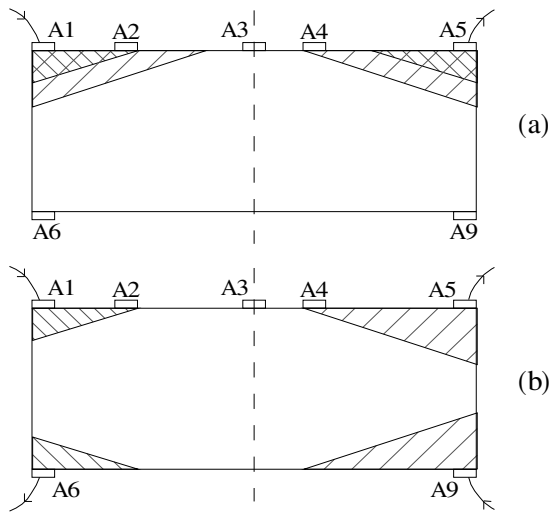}
\caption{Schematic illustration
of a lengthwise section of sample $A$ showing the propagation of resistive
fronts with current directed in the (a) $ab$ direction ($A1,A5$) and (b)
$c$ direction ($A1,A6$) or ($A5,A9$). The $c$ axis is vertical.}%
\label{fig:theofronta}%
\end{figure}

All these observations are consistent with the idea of a resistive front
moving progressively outward from the current contacts as illustrated for
sample $A$ in Fig.\ \ref{fig:theofronta}. Also, and more surprisingly, it
seems that the position of the front depends only on the magnitude of the
current injected and seems to be independent of its ultimate destination if
one is to account for the nearly identical $VI$ response on, for example, the
potential contact pair ($A2,A4$) independently of whether the current is
withdrawn on the same face ($A1,A5$) or the opposite face ($A1,A6$); similarly
for the potential pair ($A3,A4$) for current through ($A1,A5$) or ($A5,A9$).
The difference between the values of the threshold current measured on
($A2,A4$) and ($A3,A4$) contact pairs is accounted for by the distances of the
potential contact from the nearest current contact ($A2$ from $A1$ or $A4$ from
$A5$). In these experiments, the fronts did not attain contact $A3$ which
remained at the potential of the resistanceless region. The triangular form used 
for the resistive region in the figure is purely schematic to show the 
progression of the front. Further consideration of possible forms is given below 
where it is argued that the form is in fact most probably rectangular.

\begin{figure}[ptb]
\includegraphics[width=8.6cm]{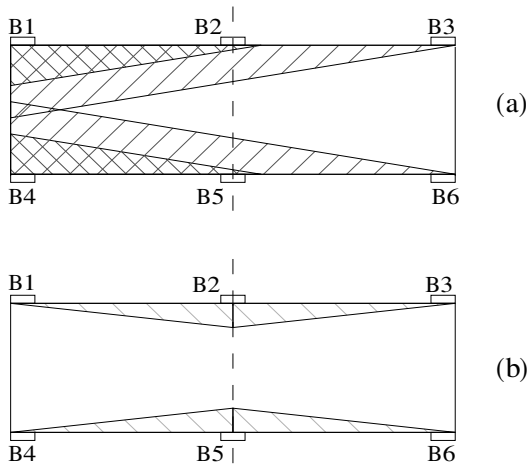}
\caption{Schematic illustration
of the propagation of resistive fronts with a $c$-directed current in a
lengthwise section of sample $B$. The current was driven through (a) ($B1,B4$)
contacts near one end of the sample (b) ($B2,B5$) at the middle of the
sample.}%
\label{fig:theofrontb}%
\end{figure}

Figure \ref{fig:theofrontb} represents how the resistive front is imagined to
progress in the experiment with sample $B$ when the current is run through the
$c$ direction. Here too the lower threshold current was obtained for potential
pair ($B2,B5$) nearest the current contacts. The resistive front from current
contacts ($B1,B4$) attains ($B2,B5$) before ($B3,B6$). When the current was
run through the middle of the sample by ($B2,B5$) the injected current is
divided between two directions with a corresponding reduction in current
density so that the voltage drop appears at higher current for the same
distance and gives the same response at both ends.

\begin{figure}
\includegraphics[width=8.6cm]{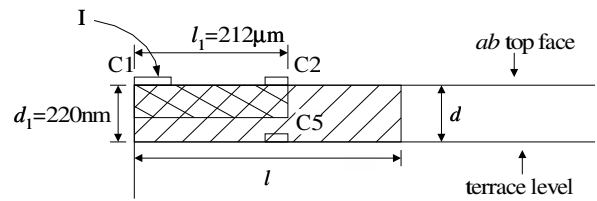}
\caption{Schematic illustration
of a section of sample $C$ with an etched terrace to investigate depth
dependence of the resistive/nonresistive front which propagates from current
contact $C1$ to attain the top surface contact $C2$ and, at higher current,
the terrace contact $C5$. If the front were rectangular as illustrated
and had a current-independent aspect ratio $\ell/d$, the dimensions would be 
linear in the total
current and $\ell/d=(\ell_{1}/d_{1})(I_{d_{1}}^{\text{th}}/I_{\ell_{1}}^{\text{th}})$
provided that $I_{d_{1}}^{\text{th}}\geqslant I_{\ell_{1}}^{\text{th}}$. }%
\label{fig:theofrontc}%
\end{figure}
Figure \ref{fig:theofrontc} represents the front for the experiment
with the terraced sample. Here the form of the front has been taken to be 
rectangular for the reasons outlined in the following section. As the front 
is defined by the line at which the current density
reaches the threshold for resistive breakdown ($ab$ plane and/or $c$ 
direction) the integral of the threshold
current density crossing the front is just the total current. Furthermore if
the front does not change geometrical form with current
amplitude its dimensions increase linearly with the current. Then, knowing the
distance from the nearest current contact and the depth of the terrace, the
threshold current values for the upper and step contacts give information on
the depth and potentially some information on the form of the front as it
attains the upper and terrace level contacts. We can expect the aspect ratio
of the front to be related to the anisotropy.

\section{Elementary Understanding}

We shall assume for the present discussion that edge effects are not
a dominant factor for our observations. Although we expect an integrable 
singularity at the edge of the sample, a large part of the middle of the 
sample has an essentially flat distribution with distance from the center line
\footnote{
Fuchs and collaborators (Ref. \onlinecite{FuchsNature}) have done a very interesting
experiment to determine the distribution of the $\mathbf{\hat{c}}$-directed
component of the current-induced magnetic field across and just outside a current carrying
sample. In the low-temperature phase which interests us here, they deduce
that there is a stronger current density on the edge of the sample. Their 
observations are compatible with the expected finite width correction 
$j_{x}=[1-(2y/w)^{2}]^{-1/2}$, where $w$ represents the full width and $y$ is measured from
the midline. This distribution is constructed to ensure the zero 
$\mathbf{\hat{c}}$-directed induced field in the sample as required by the superconducting
state with fixed vortices.} 
and we deal in the first instance uniquely with the distribution
with  depth $\mathbf{\hat{z}}$ and along the length
$\mathbf{\hat{x}}$ of the sample. We know that in the normal resistive state the current
distribution is governed by the anisotropy of the resistivity. By rescaling
the coordinates according to the anisotropy factor $\gamma_{n}=\sqrt{\rho
_{c}/\rho_{ab}}$ the problem can be solved as a Laplace equation in
the potential. The necessary coordinate rescaling is determined by the anisotropy
factor $\gamma$: $z\rightarrow\gamma z$. In the normal state, it is clearly
given by $\gamma_{n}^{2}=\rho_{c}/\rho_{ab}\approx(500)^{2}$ and
would give rise to an $ab$ current penetration depth of about $p_{ab%
}\sim\ell_{ab}/\gamma_{n}\sim$~$1$~$\mathrm{\mu m}$ for a distance
$\ell_{ab}=1$~$\mathrm{mm}$ between $ab$ current
contacts.\cite{busch1992} 

It is less recognized, perhaps, that the low current
distribution in the nonresistive superconducting regime can be governed by
the same sort of equation for the superconductor phase in similarly rescaled
coordinates if the penetration depth is smaller than the London screening
depth. In the nonresistive superconducting phase, the
Laplacian in the potential is replaced by a Laplacian in the superconductor
phase which has the same form of relationship to the superconducting current
as does the potential to the current in the normal state. The anisotropy ratio
in the superconducting state is given by $\gamma_{s}^{2}=(n_{s}e\hbar
/m_{ab})/sJ_{J}$, the ratio of the coefficients of the $ab$ and $c$
phase gradients in the relation between the current operator and the phase.
The $sJ_{J}$ term is the low current expansion of the Josephson relation
between successive superconducting planes separated by $s$, while $n_{s}$ is
the three-dimensional (3D) superconducting electron density and $1/\mathbf{m}$ the inverse
mass tensor. If one were to neglect London screening, $\gamma_{s}$ would
govern the current penetration in the same way as does $\gamma_{n}$ in the
normal state. 

The superconducting anisotropy is usually expressed as a mass
ratio or London screening length ratio for $c$ and $ab$ currents $\gamma
_{s}^{2}=m_{c}/m_{ab}=\lambda_{c}^{2}/\lambda_{ab}^{2}$ and
is customarily expected to have about the same value as the extrapolated
normal-state resistivity ratio. In BSCCO, however,
experiments\cite{latyshev1999} on mesa structures have shown a Josephson
critical current reduced by a factor of about $30$ with respect to the usual
Ambegaokar-Baratoff\cite{ambegaokar} relation to the high current resistance;
the low current resistive slope is proportionately altered to $V_{\mathrm{gap}%
}/sJ_{J}\approx30\rho_{c}$ and the coefficient in the Josephson relation
$J_{z}=J_{J}\sin(\varphi_{n}-\varphi_{n+1})\rightarrow sJ_{J}\partial
\varphi/\partial z$ at low current is reduced by the same factor with a
corresponding enhancement of the anisotropy factor [$\varphi_{n}$ is the phase
$\varphi(\mathbf{r})$ of the superconductor wave function at the $n{\mathrm{th}}$ superconducting plane]. This has been attributed to the $d$-wave nature of
the order parameter and already enhances the anisotropy factor by $\sqrt{30}$ over the normal state. But if one also considers the effect of phase
fluctuations across the Josephson junction caused by misalignment of vortices
in magnetic field, another multiplicative factor $\langle\cos\varphi
_{n,n+1}\rangle^{-1/2}$ should be incorporated into $\gamma_{s}$. The factor
$\langle\cos\varphi_{n,n+1}\rangle$ has been variously estimated from
Josephson plasma resonance \cite{matsuda1995} experiments to be between about
$10^{-3}$ and $10^{-1} $, extrapolating to about $4\times10^{-3}$
(Refs.\ \onlinecite{shibauchi1999} and \onlinecite{gaifullin2000}) at low
temperature and $\mu_{0}H=1.5$~T, which suggests a very 
strongly enhanced anisotropy parameter in the superconducting phase as
compared with the normal state: $3\times10^{4}<\gamma_{s}<1\times10^{6}$.
Current penetration in the superconducting phase is governed by a combination
of London screening and anisotropy, both of which act to limit the
penetration. But, even before including the effects of London screening, the
anisotropy limits the penetration depth $p_{ab}\sim\ell
_{ab}/\gamma_{s}$ in the middle of a 1-mm-long BSCCO sample to
$2<p_{ab}<20\,\mathrm{nm}\ll\lambda_{ab}\simeq200$~nm. Thus
in short samples of BSCCO like ours the current penetration in the
superconducting phase is always dominated by anisotropy rather than by London screening.

This suggests that the effective thickness for current flow might be given
approximately by $p_{ab}\sim\ell_{ab}/\gamma_{s}%
<\lambda_{ab}\ll t$ (the real thickness), and furthermore that the
current density is nonuniform along the sample, being higher closer to the
contacts, causing breakdown to resistive behavior to be progressive and not
to penetrate immediately throughout the bulk. Under these conditions,
$ab$ breakdown alone is not possible: if only $ab$ breakdown were to occur, the
front would have to be parallel to the $c$ direction because no part of the
interface with the nonresistive region can be parallel to the electric field
produced along the $ab$ breakdown direction. But if breakdown does not occur
uniformly throughout the thickness of the sample, the front must somewhere
change orientation. The electric field in the resistive portion, which must be
oriented perpendicular to the interface, therefore also changes orientation
and that requires $c$ breakdown. Hence a resistive/nonresistive front within
the sample which does not go straight through it along a principal direction
necessarily implies simultaneous $ab$ and $c$ breakdown. The regions of $ab$
and $c$ breakdown need not be spatially coincident however, but they must be
at least contiguous. The same can be argued for the case of pure $c$ breakdown.

Another consequence of the nonuniformity of the current distribution in depth
is the presence of shear forces on the vortices and the possibility of sliding
between planes of (semi)ordered vortex segments.\cite{busch1992,
khaykovich2000} This has important consequences on the locality of the 
$\mathbf{E}(\mathbf{J})$ relation. 
If the force required to shear two adjoining planes derives
from an effective Josephson coupling $J_{J}\langle\cos\varphi_{n,n+1}\rangle$,
the condition for nonelastic shear (slip) is $\partial J_{ab%
}/\partial z\gtrsim J_{J}\langle\cos\varphi_{n,n+1}\rangle a/s^{2},$ where $a$
and $s$ are the distances between vortices in the plane and the distance
between neighboring planes, respectively. Using the expression for the
anisotropy penetration depth $p_{ab}\sim\ell_{ab}%
/\gamma_{s}$ and identifying the effective Josephson coupling with the
$c$-axis critical current density $J_{c}^{\mathrm{th}}$, the condition for shear
to occur at the onset of dissipation is $J_{ab}^{\mathrm{th}}%
/J_{c}^{\mathrm{th}}\gtrsim\gamma a\ell_{ab}/s^{2}$. At high values
of anisotropy, $\gamma_{s}\gtrsim10^{2}$, however, the shear strength is
dominated by magnetic coupling between vortex segments.\cite{buzdinfeinberg,
koshelevkes} Either by relating the critical force per vortex segment for
shear slip $f_{c}$ to the tilt modulus $C_{44}$ by $f_{c}\approx C_{44}as$ and
using the evaluations of the modulus for magnetic coupling,\cite{koshelevkes}
or using directly the calculation of the shear strength,\cite{pe1997} it
appears that shear slip between planes should occur for $\partial
J_{ab}/\partial z\gtrsim(c\phi_{0}/32\pi^{3}\lambda_{ab}%
^{4})\langle\cos\varphi_{n,n+1}\rangle a/s$. It will be seen that this
condition is met with the values obtained from the data analyzed according to
the reasoning outlined below and it is a necessary condition to have a local
$\mathbf{E}(\mathbf{J})$ relation.

The above considerations lead to a scenario where the current injected and
withdrawn from the same $ab$ surface is confined by anisotropy to a
penetration depth $\sim\ell_{ab}/\gamma_{s}$ in the nonresistive
regime with consequent enhancement of the current density. When the current
density attains the critical value, for either $J_{c}$ or $J_{ab}$,
the current distribution is modified until both reach their breakdown values
and the sample is resistive on one side and nonresistive on the other side of
a front which advances out from each current contact point as the current is
increased. As the fronts attain successive voltage contacts, these display a
potential drop with respect to the portion of the sample which has remained in
the nonresistive state. The shape of the resistive region can be expected to be a function of the $\mathbf{E}$-$\mathbf{J}$ response around the breakdown values (vortex
depinning for $ab$ current and Josephson-junction behavior for $c$ current)
described for the most part by the anisotropy factor $\gamma$ of the resistive
state and $\Gamma=J_{ab}^{\mathrm{th}}/J_{c}^{\mathrm{th}}$. If the
shape were determined uniquely by the $\mathbf{E}$-$\mathbf{J}$ response, its position would
depend only on the contact from which it originated and the magnitude of the
current injected.
\begin{figure}
\includegraphics[width=8.6cm]{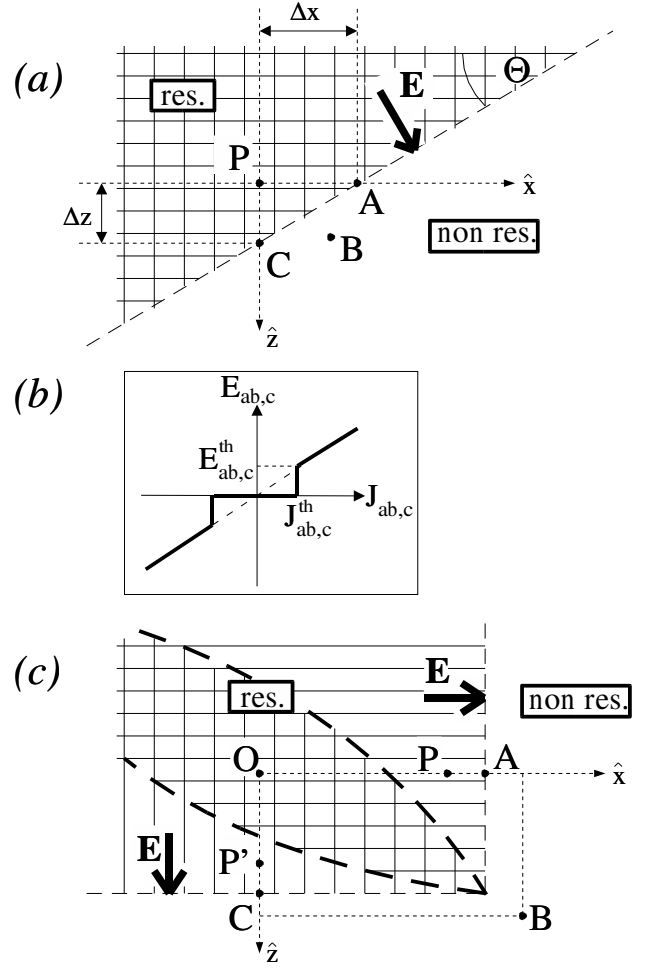}
\vskip -0.3cm
\caption{The upper part (a) of the
figure illustrates the condition that the electric field be perpendicular to
any segment of the resistive/nonresistive interface supposing spatially
coincident $ab$ and $c$ current breakdown; ``{\textbf{res}}'' denotes the resistive
side, ``\textbf{nonres}'' the nonresistive side. $E_{ab}$ and $E_{c}$ are given by
the $\mathbf{E}(\mathbf{J})$ relationship, an example of which is shown in
part (b). Positive differential resistivities for both components imply
both $\partial J_{ab}/\partial x_{ab}>0$ and $\partial J_{c}/\partial x_{c}%
>0$, in contradiction with current conservation which requires that these two
derivatives have opposite signs (for a 2D distribution). If however one of the
two current-density components is subcritical, $\Theta=0,\pi/2$, the
corresponding derivative is free. The latter situation is illustrated in the
lower part of the figure where in addition the front changes orientation from
$\Theta=0$ to $\Theta=\pi/2$ within the sample, a situation which is still
compatible with the irrotationality condition $\oint\mathbf{E}\hspace{-0.3mm}\cdot\hspace{-0.3mm}\mathrm{d}%
${\boldmath$\mathbf{\ell}$} $=0$ on the electric field provided that the
resistive region between pure $ab$ and pure $c$ breakdown also terminates at
the corner of the rectangle.}%
\label{fig:angle}%
\end{figure}

Restrictions are imposed on the shape of the front by the irrotational nature
of the electrochemical field and by current conservation. The inductive field
from the current is negligible compared to resistive fields for our pulses and
sample sizes and we can suppose the field $\mathbf{E}$ to be irrotational: $\nabla
\times\mathbf{E}=-\partial\mathbf{B}/\partial t=0.$ We can then, as in
London's treatment of breakdown in a type-I current carrying wire,\cite{Londonbook} demand that
$\oint\mathbf{E}\hspace{-0.5mm}\cdot\hspace{-0.5mm}\mathrm{d}${\boldmath$\mathbf{\ell}$} $=0$
be satisfied through the resistive/nonresistive interface
(Fig.\ \ref{fig:angle}) with the consequence that the interface must be
perpendicular to the field. As argued above, our quasipoint contact situation
enforces that both $ab$ and $c$ breakdown appear simultaneously and the $VI$
response in either configuration will contain both $c$ and $ab$ features.
Either $ab$ and $c$ breakdown occur together at the interface or they occur
alone in separated but necessarily contiguous or overlapping regions. The
first case is considered in part (a) of Fig.\ \ref{fig:angle} where a segment
of the interface is locally oriented at an angle $\Theta$ with the $ab$ plane.
A model with a local (locality implicitly assumes shearing of the vortex
segments from plane to plane) $\mathbf{E}$-$\mathbf{J} $ relation for
$\mathbf{J}>\mathbf{J}^{\mathrm{th}}$ of the form
\begin{equation*}
E_{x}=0,\qquad |J_{x}|\leqslant J_{ab}^{\mathrm{th}}
\end{equation*}
\begin{equation}
E_{x}=\pm E_{ab}^{\mathrm{th}}+\rho_{ab}^{\prime}(J_{x}\mp
J_{ab}^{\mathrm{th}}),\qquad J_{x}\gtrless\pm J_{ab}^{\mathrm{th}}
\end{equation}
\begin{equation*}
E_{z}=0,\qquad |J_{z}|\leqslant J_{c}^{\mathrm{th}}
\end{equation*}
\begin{equation}
E_{z}=\pm E_{c}^{\mathrm{th}}+\rho_{c}^{\prime}(J_{z}\mp J_{c}^{\mathrm{th}%
}),\qquad J_{z}\gtrless\pm J_{c}^{\mathrm{th}}
\end{equation}
on the resistive side is shown in part (b) of the figure. $\rho_{ab}^{\prime}$
 and $\rho_{c}^{\prime}$ are the dynamic resistivities in the $ab$
and $c$ direction in the superconductive state beyond breakdown. The condition
that the field be perpendicular to the interface would demand that $\tan
\Theta=\Delta z/\Delta x=\pm E_{ab}^{\mathrm{th}}/E_{c}%
^{\mathrm{th}}$, where the $E_{ab}^{\mathrm{th}}$ and
$E_{c}^{\mathrm{th}}$ components of the electrochemical field both point
towards the interface, in the directions of the $ab$ and $c$ components of the
current flow. However because both components of the current density must
diminish on approaching the interface, $\partial J_{x}/\partial x$ and
$\partial J_{y}/\partial y$ both have the same sign whereas current
conservation $\nabla\hspace{-0.5mm} \cdot \hspace{-0.5mm} {\mathrm{{\bf{J}}=0}}$ requires them to have opposite signs. More
explicitly, if the interface is imagined to have positive intercepts with the
axes of an ($\widehat{x},\widehat{z}$) coordinate system whose origin is at
$P$ in the resistive region as in Fig. \ref{fig:angle}, both $J_{x}%
>J_{ab}^{\mathrm{th}}$ and $J_{z}>J_{c}^{\mathrm{th}}$ (all with
positive values, current flowing from left to right and top to bottom). But
the fact that the current diminishes on approaching the interface along each of
the principal directions to reach $J_{ab}^{\mathrm{th}}$ and
$J_{c}^{\mathrm{th}}$ which defines the interface imposes that both $\partial
J_{x}/\partial x<0$ and $\partial J_{z}/\partial z<0$ at the same point $P$,
in violation of current conservation which demands that $\partial
J_{x}/\partial x=-\partial J_{z}/\partial z$. 

We conclude that a region of
spatially coincident $ab$ and $c$ breakdown may not have an interface with the
nonresistive region.

\begin{figure}[ptb]
\begin{center}
\includegraphics[width=8.6cm]{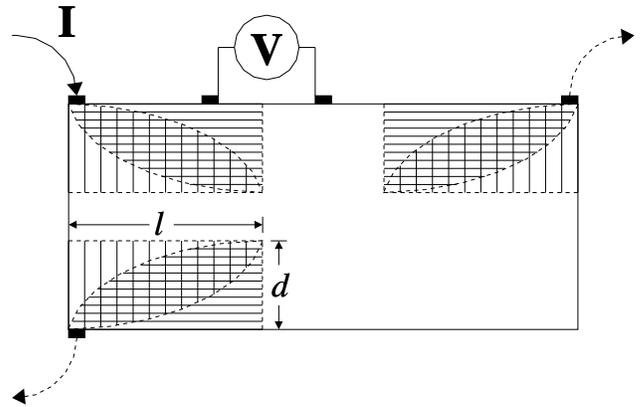}
\caption{Generic form of resistive regions. The dotted current withdrawal
arrows are $ab$ and $c$ configuration alternatives. Horizontal lines indicate
$ab$ breakdown, vertical lines $c$ breakdown.}
\label{breakdown rectangle}
\end{center}
\end{figure}

Resistive/nonresistive interfaces perpendicular to pure $ab$ or pure $c$
breakdown, however, do not violate current conservation. For sole $c$
breakdown, $J_{z}>J_{c}^{\mathrm{th}}$, $J_{x}<J_{ab}^{\mathrm{th}}$ at
the interface and the positive $\partial J_{x}/\partial x$ which must result
from the negative $\partial J_{z}/\partial z$ to conserve
current has no consequence for the electric field because the $x$ component is
subcritical and produces no electric
field. A similar argument can be made for pure $ab$ breakdown, establishing
that nonresistive/resistive interfaces are possible for single-component
breakdown whereas they are not for spatially coincident two-component breakdown. 
Furthermore it is possible to change the orientation of the
interface between the two perpendicular principal directions of the
superconductor on the condition that the two separated mutually
perpendicular $ab$ and $c$ fronts meet at
a common point and that the interface between the two types of breakdown
converges to the same point. The latter interface could be either a simple
common boundary or more generally a lens shaped region of coincident $ab$ and
$c$ breakdown. Such a situation is illustrated in part (c) of Fig.
\ref{fig:angle}. We conclude that the fronts must be formed of segments of
interfaces perpendicular to pure $ab$ or pure $c$ current breakdown. Numerical
solutions of the model\cite{Kriza} indicate a simple rectangle with a generic form of the
type illustrated in Fig. \ref{breakdown rectangle}.%

The length $\ell$ and depth $d$ of the resistive region are related to the
current by
\begin{equation}
I=w(dJ_{ab}^{\mathrm{th}}+\ell J_{c}^{\mathrm{th}}),
\label{current-interface}
\end{equation}
where $w$ is the width of the sample. The threshold current measured for the
potential between the nonresistive region and a contact on the surface at a
distance $\ell_{1}$ from the current injection contact is given by the
condition that the front arrives at the contact, $\ell=\ell_{1}$:
\[
I_{\ell_{1}}^{\mathrm{th}}=w\ell_{1}(J_{ab}^{\mathrm{th}}d/\ell+J_{c}%
^{\mathrm{th}})
\]
and for a voltage contact at depth $d_{1}$ by $d=d_{1}$:
\[
I_{d_{1}}^{\mathrm{th}}=wd_{1}(\ell/d)(J_{ab}^{\mathrm{th}}d/\ell
+J_{c}^{\mathrm{th}}),
\]
provided that $\ell(I)$ is sufficient that the front attains the contact in the
$\widehat{x}$ direction ($I_{\ell_{1}}^{\mathrm{th}}>I_{d_{1}}^{\mathrm{th}}$). In general, $\ell=\ell(I)$ and $d=d(I)$, but if there
are no other relevant lengths in the problem (distances between $ab$
configuration current contacts $\ell_{ab}$ $\gg\ell$ and $c$ configuration
current contacts $\ell_{c}=t\gg d$) one can expect the ratio $\ell/d$ to be
independent of the current and only a function of the anisotropy factors
$\gamma$ and $\Gamma=J_{ab}^{\mathrm{th}}/J_{c}^{\mathrm{th}}$. If
we interpret the top and terrace thresholds of the experiment on sample $C$ as
measurements of $I_{\ell_{1}}^{\mathrm{th}}$ and \ $I_{d_{1}}^{\mathrm{th}}$ and we make the
hypothesis that the aspect ratio $\ell/d$ is independent of current, then the
measurements tell us that
\[
\ell/d=(\ell_{1}/d_{1})(I_{d_{1}}^{\mathrm{th}}/I_{\ell_{1}}^{\mathrm{th}})\approx2500
\]
and
\begin{equation}
J_{ab}^{\mathrm{th}}d/\ell+J_{c}^{\mathrm{th}}\approx10\ \mathrm{A}\,\mathrm{cm}^{-2}
\label{limit Jc}
\end{equation}
or, multiplying by $\ell/d$,
\begin{equation}
J_{ab}^{\mathrm{th}}+J_{c}^{\mathrm{th}}\ell/d\approx2.5\times10^{4}\ \mathrm{A}\,\mathrm{cm}^{-2},
\label{limit Jab}
\end{equation}
where the last two quantities are the same combination of $ab$ and $c$
transport properties which we cannot separate without a model calculation for
the aspect ratio of the resistive front. This will be the subject of a future
paper\cite{Kriza} on numerical solutions of the present model.

The upper limits on $J_{c}^{\mathrm{th}}$ and $J_{ab}^{\mathrm{th}}$
given by relations (\ref{limit Jc}) and (\ref{limit Jab}) are consistent with the
value $J_{c}^{\mathrm{th}}\approx2$~A\,cm$^{-2}$ obtained from combining $J_{J}$
measured in the mesa experiment with $\langle\cos\varphi_{n,n+1}\rangle
\approx4\times10^{-3}$ extrapolated from the Josephson plasma resonance
experiments \cite{shibauchi1999} and the value $J_{ab}^{\mathrm{th}}\approx2\times10^{4}$ A\,cm$^{-2}$ estimated from the magnetic hysteresis.\cite{khaykovich2000} The
condition for slip between planes is well satisfied for these values
indicating that the description is self-consistent.

\section{Summary}

The experimental results on $VI$ characteristics for $ab$ surface contacted
samples of the highly anisotropic BSCCO superconductor in the low-temperature
vortex lattice phase above about $2000$~Oe show that the $VI$ responses for
$ab$ or $c$ current configurations are virtually identical and that the
dissipative region invades the samples in the form of a
resistive/nonresistive front moving out from the current contacts with
increasing current. The $VI$ characteristic depends only on current
magnitude, voltage contact position and nearest current injection point. It is
argued that current penetration in the nonresistive regime is limited by the
high anisotropy to depths less than the London screening length, and that the
resistive region must involve simultaneous $ab$ and $c$ breakdown and shear of
successive planes of vortex segments. The shape of the resistive region is
argued to be composed of rectangular segments and seems to be a simple
rectangle with interfaces to the nonresistive portion consisting of pure $ab$
or pure $c$ breakdown. A different contact arrangement using an ion etched terrace
to sample the potential in the depth of the sample brings extra information
that allows access to the aspect ratio of the rectangle. But because the $ab$
and $c$ configuration experiments measure the same combination of $ab$ and $c$
properties, it is necessary to propose and solve a specific model to separate
them. A detailed discussion of numerical solutions of the model has been
reserved for future publication.

\begin{acknowledgments}
We take pleasure in acknowledging discussion with I.~T\"utt\H{o} and
L.~F.~Kiss. We are grateful to L.~Forr\'o and to B.~Keszei for supplying the
crystals used in this study. B.S. and \'A.P. acknowledge the Atomic Energy
Commission for a Bursary to visit the Saclay laboratory. Research in Hungary
has been supported by Grant Nos. OTKA T037976 and TS040878. We are grateful also for the
support of the ``BALATON'' Collaboration program of the French and Hungarian
Foreign Affairs Ministries.
\end{acknowledgments}

\bibliography{current}

\end{document}